# Sound transpilation from binary to machine-independent code

Roberto Metere[1], Andreas Lindner[2], and Roberto Guanciale[2]

[1] Newcastle University, UK,
    r.metere2@ncl.ac.uk
[2] KTH Royal Institute of Technology, Sweden,
    {andili,robertog}@kth.se

**Abstract.** In order to handle the complexity and heterogeneity of modern instruction set architectures, analysis platforms share a common design, the adoption of hardware-independent intermediate representations. The usage of these platforms to verify systems down to binary-level is appealing due to the high degree of automation they provide. However, it introduces the need for trusting the correctness of the translation from binary code to intermediate language. Achieving a high degree of trust is challenging since this transpilation must handle (i) all the side effects of the instructions, (ii) multiple instruction encoding (e.g. ARM Thumb), and (iii) variable instruction length (e.g. Intel). We overcome these problems by formally modeling one of such intermediate languages in the interactive theorem prover HOL4 and by implementing a proof-producing transpiler. This tool translates ARMv8 programs to the intermediate language and generates a HOL4 proof that demonstrates the correctness of the translation in the form of a simulation theorem. We also show how the transpiler theorems can be used to transfer properties verified on the intermediate language to the binary code.

## 1 Introduction

Despite the existence of formally verified compilers, the verification of binary code is a critical task to guarantee trustworthiness of critical systems. This is particularly necessary for software mixing high-level language with assembly (system software), using ad-hoc languages and compilers (specialized software), in presence of instruction set extensions (like encryption and decryption), and when the source code is not available (binary blobs). This necessity is not only limited to the general-purpose computing scenario but also applies to connected embedded systems, where software bugs can enable a remote attacker to tamper with the security of automobiles, payment services, and smart IoT devices.

The need of semi-automatic analysis techniques for binary code has lead to the development of several tools [25,7,24]. To handle the complexity and heterogeneity of modern instruction set architectures (ISAs), all these tools followed a common design: They have introduced a platform independent intermediate representation that allows to implement analysis independently of (i) names and number of registers, (ii) instruction decoding, (iii) endianness of memory access, and (iv) instruction side-effects (like updating conditional flags or the stack pointer). This intermediate representation is often a dialect of the Valgrind's IR [21]. Soundness of the transpiler (i.e. the tool translating from machine code to intermediate language) should not be foregone: It may have to handle multiple instruction encoding (e.g. ARM Thumb), variable instruction length (e.g. Intel), and complex side effects of instructions (e.g. ARM branch with link and conditional executions). Clearly, a transpiler bug jeopardizes the soundness of all analyses done on the intermediate representation.

Our strategy to handle this issue is to use formal models of the ISA and of the intermediate language of the analyses platform, and to formally demonstrate that the transpilation is correct. We chose ARMv8 [17] as demonstrating ISA, reusing the model for the HOL4 theorem prover that was previously developed in [10,9]. For the target language, we implemented a deep-embedding of the Intermediate Language of the Binary Analysis Platform [7] (BIL) in the HOL4 logic and implemented its small-step semantics. Verification of the transpilation is done via a HOL4 proof producing transpiler, which translates ARMv8 programs to BIL programs, and yields the HOL4 proof that demonstrates its correctness. The theorem establishes a simulation between the input binary program and the generated BIL program, showing that the two programs have the same behavior. Our contribution enables a verifier to prove properties of the generated BIL program (i.e. by directly using the theorem prover or proof-producing analysis techniques) and to transfer them to the original ARMv8 program using the generated simulation theorems.

*Outline* We present the state of the art and the previous works relating to our contribution in Section 2. Section 3 introduces the HOL4 formal models of the ARMv8 ISA and the BIL language. Section 4 presents the certifying transpiler. We demonstrate that the theorems produced by the transpiler can be used to transfer verification conditions in Section 5, where we test and evaluate our development too. We give concluding remarks in Section 6.

## 2 Related work

Recent works have shown that formal techniques are ready to achieve detailed verification of real software, making it possible to provide low-level platforms with unprecedented security guarantees [13,1,8]. For such system software, limiting the verification to the source code level is undesirable. A modern compiler (e.g. GCC) consists of several millions of lines of code, in contrast to microkernels that consist of few thousand lines of code, making it difficult to trust the compiler output even when optimization is disabled[3].

---

[3] An example of a very recent bug found in GCC: https://gcc.gnu.org/bugzilla/show_bug.cgi?id=80180

To overcome this limitation, formally verified compilers [15,6,14] and proof-producing compilers [16] have been developed. Similarly to our work, these compilers use detailed models of the underlying ISA to show the correctness of their output. This usually involves a simulation theorem, which demonstrates that the behavior of the produced binary code resembles the one specified by the semantics of the high level language (e.g. C or ML). These theorems permit properties verified at the source-level to be automatically transferred to the binary-level. For instance, CompCert has been used in [3] to verify security of OpenSSL HMAC by transferring functional correctness of the source code to the produced binary.

Even if formally verified compilers obviate the need for trusting their output, they do not fulfill all the needs of verified system software. Some of these compilers target languages that are unsuitable for developing system software (e.g. ML cannot be used to develop a microkernel due to its garbage collector). Also, they do not support mixing the high-level language with assembly code, which is necessary for storing and restoring the CPU context or for managing the page table. Some of the effects of these operations can break the assumptions made to define a precise semantics of the high level language (e.g. a memory write can alter the page table which in turn affects the virtual memory layout). Also, some properties (e.g. absence of side channels due to non-secure accesses to the caches) cannot be verified at the source code level; the analysis must be aware of the exact sequence of memory accesses performed by the software. Finally, binary blob analysis is imperative for verifying memory safety of binary code whose source code is not available (e.g. the power management of ARM trusted firmware).

Unfortunately, detailed formal specifications of machine languages (e.g. the ones used to verify compiler correctness [11]) consist of thousands of lines of definitions. The complexity of these models makes them unusable to directly verify any binary code that is not a toy example. Moreover, the target verification tools, usually interactive theorem provers, provide little or no support for either automatic reasoning or reuse of algorithms among different hardware models. To make machine-code verification proofs reusable by different architectures, Myreen et al. [20] developed a proof-producing decompilation procedure. Those tools have been implemented in the HOL4 system and have been used by the seL4 project to check that the binary code produced by the compiler is correct, permitting to transfer properties verified at the source code level to the actual binary code executed by the CPU [22]. The same framework has been used to verify a `bignum` integer library [19]. However, the automatism provided by this framework is still far from what is provided by today's binary analysis platforms (e.g. [25,7,24]). These provide tools to compute and analyze control-flow graphs, to perform abstract interpretation and symbolic execution, to verify contracts, and to verify information flow properties [2]. On the other hand, their usage requires to trust the used transpiler. Due to the complexity of writing a transpiler for each architecture, recent work has been done to synthesize the transpiler from compiler backends [12]. However, this requires to trust both: the synthesis procedure and the compiler backend.

In this paper, we address this issue by providing sound transpilation of ARMv8 binary code to the intermediate language of BAP. BAP is an analysis platform that provides utilities to compute and analyze control-flow graphs, to transform programs (e.g. by unrolling cycles), to verify contracts via generation of weakest preconditions and their export to SMT solvers. The platform has also been externally extended with tools for information flow security based on relational analysis. We developed a HOL4 formal model of the BAP intermediate language, which can be used to provide precise semantics of programs expressed in BIL and to verify soundness of analysis tools. This allows us to implement a proof-producing transpiler, which can translate an ARMv8 program to a BIL program while generating a HOL4 proof that demonstrates its correctness.

## 3  Formal HOL4 models

### 3.1  The ARMv8 model

In our work, we use the ARMv8 model developed by Fox [10], which is constructed from the pseudocode described in the ARM specification [17] and provides a detailed HOL4 formalization of the effects of the instructions, taking into account the different execution modes, flags, and other characteristics of the processor behavior.

The system state is modeled as a tuple $s = \langle r, sr, p, c, m \rangle$. Here, $r$ represents a sequence of 64-bit general purpose registers. We identify the $i$-th register with $r(i)$. The tuple $sr = \langle pc, sp, lr \rangle$ contains the special registers representing the program counter, the stack pointer, and the link register respectively. The tuple $p$ representing the current processor state and contains the arithmetical flags, the execution mode, and the interrupt disabling. The tuple $c$ encodes the system and coprocessor registers, it also contains the current endianness and the configuration of the Memory Management Unit. The 64-bit addressable memory is modeled as the function $m : \mathbb{B}^{64} \to \mathbb{B}^{8}$. Finally, the system behavior is represented by the deterministic transition relation $s \to s'$, describing how the ARM state $s$ reaches the state $s'$ by executing a single instruction. Hereafter, we use . to access tuple fields; for example $s.sr.pc$ states for the program counter of the state $s$.

The HOL4 model consists of hundreds of definitions and its complexity makes it difficult to analyze large programs. To simplify the analyses, the model is equipped with a mechanism to statically compute the effects of a single instruction via the $arm\_step$ function. Let $i$ be the binary encoding of an instruction and $ad$ be the address where the instruction is stored, then the function $arm\_step(i, ad)$ returns a list of step theorems $[st_1, \ldots, st_n]$. Each theorem $st_j$ has the following structure:

$$\forall s.\, \text{read}_{32}(s.m, s.sr.pc) = i \land s.sr.pc = ad \land c_j(s) \Rightarrow s \to t_j(s)$$

where $\text{read}_{32}$ is a function that reads 32 bits from the memory. Intuitively, each step theorem describes one of the possible behaviors of the instruction and consists of the guard condition $c_j$ that enables the transition and the function $t_j$ that transforms the starting state into the next state. We use three examples to illustrate this mechanism.

Let the instruction stored at the address `0x1000000c` be the addition of the registers $x0$ and $x1$ into the register $x0$ (whose encoding is `0x8b000020`), the step function produces the following step theorem:

$$\forall s.\, \text{read}_{32}(s.m, s.sr.pc) = \texttt{0x8b000020} \land s.sr.pc = \texttt{0x1000000c} \Rightarrow$$
$$s \to \left(\lambda s'.s' \text{ with } r(0) = s'.r(0) + s'.r(1) \text{ with } sr.pc = s'.sr.pc + 4\right) s$$

(where $s'$ with $r(0) = v$ updates the register zero of the state $s'$ with $v$). In this case, only one theorem is generated, and there is no guard condition (i.e. $c_1$ is a tautology).

Some ARMv8 instructions (i.e. conditional branches) can have different behavior according to the value of some state components. In these cases, the step function produces as many theorems as the number of possible execution cases. For example, the output of the step function for the Signed Greater Than (GT) branch consists of the following two theorems:

$$\forall s.\, \text{read}_{32}(s.m, s.sr.pc) = \texttt{0x54fffe8c} \land s.sr.pc = \texttt{0x1000000c}$$
$$\land\ s.p.Z = 0 \land s.p.N = s.p.V \Rightarrow$$
$$s \to (\lambda s'.s' \text{ with } sr.pc = s'.sr.pc - \texttt{0x30}) s$$

$$\forall s.\, \text{read}_{32}(s.m, s.sr.pc) = \texttt{0x54fffe8c} \land s.sr.pc = \texttt{0x1000000c}$$
$$\land \neg\, (s.p.Z = 0 \land s.p.N = s.p.V) \Rightarrow$$
$$s \to (\lambda s'.s' \text{ with } sr.pc = s'.sr.pc + 4) s$$

That is, if the test succeeds (i.e. $c_1 = s.p.Z = 0 \land s.p.N = s.p.V$ holds) then the jump is taken (in this case jumping back in a loop to the address $pc - \texttt{0x30}$), otherwise (i.e. $c_2 = \neg(s.p.Z = 0 \land s.p.N = s.p.V)$ holds) the jump is not taken (the program counter is updated to point to the next instruction). Notice that for every state $s$ the condition $c_1 \lor c_2$ hold.

Finally, some ARMv8 instructions (i.e. memory stores) can have unsound behavior if some conditions are not met. In these cases, the step function generates the step theorems only for the correct behaviors; for a given instruction, let $st_1, \ldots, st_n$ be the generated theorems and $c_1, \ldots, c_n$ the corresponding guards, the behavior of the instruction is soundly deduced by the step function for every state $s$ such that $\bigvee_j c_j(s)$ holds and can not be deduced otherwise. For example, the output of the step function for a memory store consists of the theorem:

$$\forall s.\, \text{read}_{32}(s.m, s.sr.pc) = \texttt{0xf90007e0} \land s.sr.pc = \texttt{0x1000000c}$$
$$\land\ aligned(s.sr.sp + 8) \Rightarrow$$
$$s \to \begin{pmatrix} \lambda s'.s' \text{ with } m = write_{64}(s'.m, s'.sr.sp\ + 8, s'.r(0)) \\ \text{with } sr.pc = s'.sr.pc + 4 \end{pmatrix} s$$

Intuitively, the step function can predict the behavior only for states having the target address (i.e. $s.sr.sp + 8$) aligned.

### 3.2 The BIL model

The target of our transpilation is BIL. In this language, a statement has only explicit state changes, i.e. there are no implicit side effects, and it can only affect one variable.

BIL's syntax is depicted in Table 1. A program is a list of blocks, each one consisting of a uniquely identifying label (i.e. a string or an integer) and a list of atomic statements. A statement can affect the state by (i) assigning the evaluation of an expression to a variable, (ii) (conditionally or unconditionally) modifying the control flow, (iii) halting the system in a successful state, and (iv) terminating the system in a failure state if an assertion does not hold. As usual, labels are used to refer to the specific locations in the program and can be the target of jump statements. BIL expressions are built using constants (i.e. strings and integers), conditionals (i.e. **ifthenelse**), standard binary and unary operators (ranged over by $\Diamond_b$ and $\Diamond_u$ respectively) for finite integer arithmetic, and accessing variables of the environment (i.e. **var**). Additionally, two types of expressions can operate on memories. The expression **load**$(exp_1, exp_2, \tau_{reg,n})$ reads $n$ bytes from the memory $exp_1$ starting from the address $exp_2$. The expression **store**$(exp_1, exp_2, exp_3, \tau_{reg,n})$ returns a new memory in which all the locations have the same values as the initial memory $exp_1$ except the addresses $exp_2 + i$ where $i \in [0 \ldots n-1]$ that contain the chunks of $exp_3$.

$$
\begin{aligned}
program &:= block^* \\
block &:= (string \mid integer, stmt^*) \\
stmt &:= \textbf{assign}\,(string, exp) \mid \\
&\quad\;\; \textbf{jmp}\,(exp) \mid \textbf{cjmp}\,(exp, exp, exp) \mid \\
&\quad\;\; \textbf{halt} \mid \textbf{assert}\,(exp) \\
exp &:= string \mid integer \mid \\
&\quad\;\; \textbf{ifthenelse}\,(exp, exp, exp) \mid \\
&\quad\;\; \Diamond_u\, exp \mid exp\, \Diamond_b\, exp \mid \textbf{var}\, string \mid \\
&\quad\;\; \textbf{load}\,(exp, exp, \tau) \mid \textbf{store}\,(exp, exp, exp, \tau)
\end{aligned}
$$

**Table 1.** BIL's syntax

Hereafter we use $\Delta$ to represent the set of all possible strings. These can be used to identify both labels and variable names. We use $\tau$ to range over BIL data types; let $n \in \{1, 8, 16, 32, 64\}$, the type for words of $n$-bits is denoted by $\tau_{reg,n}$ and the type for memories addressed using $n$-bits is denoted by $\tau_{mem,n}$. We use $T$ and $V$ to represent the set of all BIL types and values respectively.

A program $b$ is well-defined if it has no duplicate block labels and each block has at least one statement. In the following we assume that all programs are well defined. Notice that the program $b$ is not part of the state, since it is not allowed to be changed dynamically.

A BIL environment $\sigma$ maps variable names (given as strings) to pairs of type and value; $\sigma : \Delta \to (T \times V)$. Types of variables are immutable and any wrongly typed operation produces a run-time failure. The semantics of BIL expressions is modeled by the evaluation function $eval$: It takes an expression $\alpha$ and an environment $\sigma$ and yields either a value having a type in $T$ or $\bot$. The evaluation intuitively follows the semantics of operations by recursively evaluating the sub-expressions given as operands. The value $\bot$ results when operators and types are incompatible, thus modeling a type error, which in turn is used by the statement semantics to cause the program counter to transition to the error state $\bot$.

A BIL state $\gamma = (\sigma, p) \in \Gamma$ is a pair of an environment $\sigma$ and a program counter $p$. Let $L = \Delta \cup \mathbb{B}^{64}$ be the set of all labels, a program counter $p$ is an element of the set $\Lambda = (L \times \mathbb{N}) \cup \{\bot, \top\}$. While executing a program, the program counter is $(l, n) \in L \times \mathbb{N}$, where $l$ is the label of the executing block and $n$ is the index for the executing statement within this block. A successfully halting program results in the program counter being $\top$. Failures (e.g. type mismatch or failing assertion) terminate the program and set the program counter to $\bot$.

The system behavior is modeled by the deterministic transition relation $b : \gamma \rightsquigarrow \gamma'$, which describes the execution of one BIL statement. In HOL4, this relation is modeled by the execution function $exc$, which defines the small step semantics of one statement.

The execution of **assign**$(X, \alpha)$ assigns the evaluation of the expression $\alpha$ to the variable $X$. Let $v = eval(\alpha, \sigma)$ and $t$ be the type of $v$, the value of the variable is updated in the context $(\sigma[X \leftarrow (t, v)])$ and the program counter is incremented. The statement fails in case of a type mismatch: $v = \bot$ or $\sigma(X) = (t', \_) \land t \neq t'$.

The statement **halt** sets the program counter to $\top$ and thereby terminates execution. The statement **assert**$(\alpha)$ just increments the program counter if the expression evaluates to true (i.e. $(\tau_{reg,1}, 1) = eval(\alpha, \sigma)$) and terminates in an error state otherwise.

The execution of **jmp**$(\alpha)$ jumps to the beginning of the referenced block, by setting the program counter to $(eval(\alpha, \sigma), 0)$. If the type of $\alpha$ is neither string nor integer then the statement fails. The statement **cjmp**$(\alpha_c, \alpha_1, \alpha_2)$ changes the control flow based on the condition $\alpha_c$. The statement fails if the type of the condition is not $\tau_{reg,1}$ or the the targets (i.e. $eval(\alpha_1, \sigma)$ or $eval(\alpha_2, \sigma)$) are not valid labels. Notice that the targets of the jump are evaluated using the current context, allowing BIL to express indirect jumps that are resolved at run-time.

## 4  The transpiler

The translation procedure uses a mapping of HOL4 ARM states to BIL states. Every ARM state field is mapped to a BIL variable or to the program counter: For example, the variable $R0$ represents the register number zero, the variable $MEM$ represents the system memory, and the BIL program counter reflects the

ARM program counter. This mapping induces a simulation relation $\sim \subseteq \Gamma \times S$ that relates BIL states to ARM states.

To transform an ARM program to the corresponding BIL fragment we need to capture all the possible effects of the program execution in terms of affected registers, flags and memory locations. The generated BIL fragment should emulate the behaviour of the instructions executed on an ARM machine. This goal is accomplished by reusing the *arm_step* function and the following three HOL4 certifying procedures.

- A procedure to translate HOL4 word terms (i.e. those having type $\mathbb{B}^{64}$, $\mathbb{B}^8$, $\mathbb{B}$ etc.) to BIL expressions. This procedure is used to convert the guards of the step theorems and the expressions contained in the transformation functions.
- A procedure to translate a single instruction to the corresponding BIL fragment. This procedure computes the possible effects of an instruction using the transformation functions of the step theorems. It also symbolically executes the resulting BIL fragment to demonstrate that it emulates the effects of the translated instruction.
- A procedure that glues together the theorems produced for the instructions to translate the entire ARM program.

To phrase the theorem produced by the transpiler we introduce the following notations. An ARM program $\pi$ is represented by a finite set of pairs $(ad_j, i_j)$, where each pair represents that the instruction $i_j$ is located at the address $ad_j$. The predicate $stored(s, \pi)$ states that the program $\pi$ is stored in the memory of the state $s$ (formally, $stored(s, \pi) \stackrel{\text{def}}{=} \forall (ad_j, i_j) \in \pi.\, \text{read}_{32}(s.m, ad_j) = i_j$). The predicate $start\text{-}block(p)$ holds if a BIL program counter $p$ points to the first statement of a block. For readability, let $\gamma = (\sigma, p)$, we use $\gamma \neq \bot$ and $start\text{-}block(\gamma)$ to denote $p \neq \bot$ and $start\text{-}block(p)$ respectively. We denote $n$ transitions of ARM states with $\rightarrow^n$, and $n$ transitions of BIL states with $\leadsto^n$. The translation procedure produces a theorem that resembles compiler correctness[4]:

**Theorem 1.** Let $ad_0$ be the entry point of the ARM program $\pi$. For every ARM state $s$ and BIL state $\gamma$, if $stored(s, \pi)$, $s.sr.pc = ad_0$, and $\gamma \sim s$, then

1. for every $n > 0$ if $s \rightarrow^n s'$ then
    $\exists n' > 0.\, b : \gamma \leadsto^{n'} \gamma' \wedge (\gamma' = \bot \vee \gamma' \sim s')$, and
2. for every $n' > 0$ if $b : \gamma \leadsto^{n'} \gamma' \wedge start\text{-}block(\gamma') \wedge \gamma' \neq \bot$ then
    $\exists n > 0.\, s \rightarrow^n s' \wedge \gamma' \sim s'$.

The meaning of the transpiler theorem is depicted in Fig. 1a. Each ARM instruction is translated to a single BIL block consisting of multiple statements.

---

[4] The ARM and BIL transition systems are deterministic and live, thus the transition relations are total functions. For this reason we omit quantifiers over the states on the right hand side of transitions, since they always exist and are unique.

Assuming that the program is stored in the ARM memory, the state is configured to start the execution from the entry point $ad_0$ of the program, and the initial HOL4 ARM state resembles the initial BIL states, then (1) for every state $s'$ reachable by the ARM model, there is an execution of the BIL program $b$ that results (after $n'$ statements) in either an error state ($\gamma' = \bot$) or in a state $\gamma'$ that resembles $s'$, and (2) for every state $\gamma'$ reachable by the BIL program after the competition of a block ($\textit{start-block}(\gamma')$), there is an execution of the ARM program that re-establishes the simulation relation.

Error states permit to identify if an initial configuration can cause a program to reach a state that cannot be handled by the transpiler (e.g. self-modifying programs or programs containing instructions whose behavior can not be predicted by the step function). It is worth noticing that these cases can not be identified statically without knowing the program preconditions (e.g. misaligned memory accesses can be caused by the initial content of the stack where pointers are stored).

### 4.1 Translation of expressions

In order to build the transpiler on top of the step function, the HOL4 expressions occurring in the guards and the transformation functions must be converted to BIL expressions. For example, while translating the binary instruction 0x54fffe8c of Section 3.1 to a conditional jump, the expressions $s.p.\text{Z} = 0 \land s.p.\text{N} = s.p.\text{V}$ and $s'.sr.pc - \text{0x30}$ must be expressed in BIL to generate the condition and the target of the jump respectively.

Let $e$ be a HOL4 expression, the output of the transpiler is the theorem $\forall \sigma. A(\sigma) \Rightarrow (eval\,(\alpha, \sigma) = e)$, stating that, if the environment satisfies the assumption $A$, then the evaluation of $\alpha$ is $e$. These assumptions usually constrain the values of the variables in the environment to match the free variables of the HOL4 expressions. For instance, for the expression $s.p.\text{N} = s.p.\text{V}$ the transpiler generates the theorem $\forall \sigma, s.(\sigma(''N'') = (\tau_1, s.p.\text{N}) \land \sigma(''V'') = (\tau_1, \text{V})) \Rightarrow (eval\,((\textbf{var}\,''N'' = \textbf{var}\,''V''), \sigma) = (\text{N} = \text{V}))$.

If a HOL4 operator has no direct correspondence in BIL, the transpiler uses a set of manually verified theorems to justify the emulation of the operator via a composition of the primitive BIL operators. This is the case for expressions that involve conversion of words to natural numbers and arithmetic operations with arbitrary precision. A relevant example is the computation of the *carry (overflow) flag* in 64-bit additions. Following the pseudocode of the ARMv8 reference manual [17], the step theorem contains the expression $[x] + [y] < 2^{64}$, where $x, y \in \mathbb{B}^{64}$ and $[\cdot] : \mathbb{B}^{64} \to \mathbb{N}$ is their interpretation as natural numbers. Both the inequality and the addition cannot be directly converted as BIL expression, because BIL can only handle numbers up to 64 bits. For the *carry flag* the transpiler uses the theorem $\forall n > 0.\ \forall x, y \in \mathbb{B}^n.\ ([x] + [y] < 2^n) \Leftrightarrow (x \gg 2 + y \gg 2 + (x\ \&\ 1) * (y\ \&\ 1) < 2^{n-1})$.

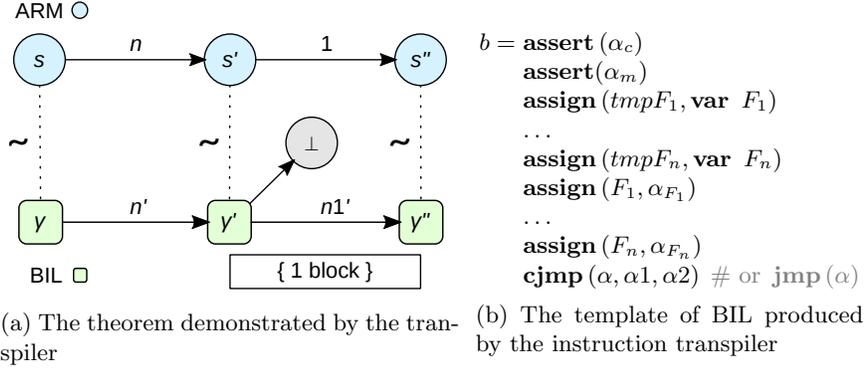

(a) The theorem demonstrated by the transpiler

(b) The template of BIL produced by the instruction transpiler

**Fig. 1.** Translating procedure

### 4.2 Translation of single instructions

The transpilation of a single instruction takes three arguments: the binary code $i$ of the instruction, the address $ad$ of the instruction in memory, and a HOL4 predicate $q_m : \mathbb{B}^{64} \to \mathbb{B}$. The latter argument identifies which memory addresses should not be modified by the instruction and is used to guarantee that the ARM program is not self-modifying. In fact, a self-modifying program cannot be transformed to equivalent BIL programs (due to BIL following the Harvard architecture). If an instruction modifies the program code then then the translated BIL program must terminate in an error state. The predicate $q_m$ is used to instrument the instruction transpiler with the information about where the program code is stored.

An ARM instruction is translated to a single BIL block, following the template of Fig. 1b. Hereafter we detail its generation and the verification of its correctness.

The transpiler uses the *arm_step* function to compute the behavior of the input instruction $i$ and to generate the step theorems $[st_1, \ldots, st_n]$. These are used to demonstrate $\forall s.(\text{read}_{32}(s.m, s.sr.pc) = i \wedge s.sr.pc = ad) \Rightarrow s \to t(s)$ where $t(s) = if\ c_1(s)\ then\ t_1(s)\ else\ if\ \ldots\ else\ if\ c_n(s)\ then\ t_n(s)$, and $c_j$ and $t_j$ are the guards and transformation functions of the step theorems respectively.

The behavior of the instruction can be soundly deduced by the step function only if one of the $c_j$ predicate holds (see Section 3.1). The transpiler simplifies the disjunction of the guards demonstrating $\forall s. \bigvee_j c_j(s) = e_c$ ( where $e_c$ is a HOL4 predicate) and translates it to a BIL expression $\alpha_c$ ( demonstrating $\forall \sigma, s.((\sigma, p) \sim s) \Rightarrow (\text{eval}\,(\alpha_c, \sigma) = e_c)$ ). The BIL statement **assert**$(\alpha_c)$ is generated as preamble of the instruction. Intuitively, if an ARM state $s$ does not satisfy any guard, then any similar BIL state $(\sigma, p)$ does not satisfy the assertion, causing the BIL program to terminate in a error state. On the other hand, if the BIL state satisfies the assertion, then every similar ARM state satisfies at least one of the guards, thus the instruction's behavior can be deduced by the step function.

The second task is to translate the effects of the instruction on every field of the ARM state. Let $f$ be one field of the ARM state (e.g. $f = r(0)$ is the register zero) and let $F$ be the corresponding variable of BIL according to the relation $\sim$. The transpiler uses HOL4 rewriting to compute the new value $e_F$ of the field (and demonstrating $\forall s.(t(s)).f = e_F$). If $e_F = s.f$ then the ARM field is not affected by the instruction and the corresponding variable $F$ should not be modified by the generated BIL block, otherwise the variable $F$ must be updated accordingly. The expression $e_F$ is translated to obtain the theorem $\forall \sigma.eval\,(\alpha_F, \sigma) = e_F$ and the BIL statement **assign**$(F, \alpha_F)$ is generated. A complication raises when there are instructions affecting several state variables, and whose resulting values depend on each other (i.e. imagine an instruction swapping registers zero and one, where $t(s) = s$ with $\{r(0) = s.r(1)\ and\ r(1) = s.r(0)\}$). To handle these cases, the translation procedure generates a statement **assign**$(tmpF, \mathbf{var}\ F)$, which backups the value of the variable $F$ into the temporary variable $tmpF$.

Special care is needed for memory updates (i.e. $f = m$). The BIL program should fail if it updates a memory location where $q_m$ holds. The transpiler inspects the expression $e_{MEM}$ to identify the addresses that can be changed by the instruction and extracts the corresponding set of $\mathbb{B}^{64}$ expressions $e_1, \ldots, e_n$ (in ARM a single instruction can store multiple registers). To ensure that this identification is complete, the transpiler proves $\forall s, a.(\bigwedge_i a \neq e_i) \Rightarrow (e_{MEM}(a) = s.m(a))$. The expression $\bigwedge_i \neg q_m(e_i)$ (which guarantees that no modified address belongs to the reserved memory region) is translated to obtain the theorem $\forall \sigma.eval\,(\alpha_m, \sigma) = \bigwedge_i \neg q_m(e_i)$. Finally the BIL statement **assert**$(\alpha_m)$ is added as further preamble of the instruction. If the ARM instruction modifies an address in $q_m$, then the corresponding BIL state does not satisfy the assertion, causing the BIL program to terminate in an error state.

Symbolic evaluation of the program counter field is used to generate statements that update the control flow. If $e_{pc}$ is syntactically equivalent to *if c then $e_1$ else $e_2$*, then the expressions $c$, $e_1$ and $e_2$ are translated to $\alpha_c$, $\alpha_1$ and $\alpha_2$, and the statement **cjmp**$(\alpha_c, \alpha_1, \alpha_2)$ is appended as last statement of the BIL fragment. Otherwise, $e_{pc}$ is directly translated to $\alpha$ and **jmp**$(\alpha)$ is appended to the BIL fragment. Whenever possible, $e_1$, $e_2$, or $e_{pc}$ are first simplified to constants, thus reducing the number of indirect jumps in the BIL program.

To compute the effects of the generated BIL block, the transpiler uses a small symbolic execution engine. The transpiler uses the intermediate theorems generated during the process to discard the hypotheses of the symbolic execution and to instantiate the expression evaluations. Finally, it establishes the *instruction-theorem*.

**Theorem 2.** Let $i$ be the binary encoding of the instruction, $ad$ be its location in memory, and $q_m$ the predicate identifying the memory region used to store the complete program. Also, let *block* be the generated BIL block, $n$ be the corresponding number of BIL statements, and $b[ad]$ be the BIL block of the BIL program $b$ having label $ad$. For every ARM state $s$, BIL state $\gamma$, and BIL program $b$ if $read_{32}(s.m, s.sr.pc) = i$, $s.pc = ad$, $\gamma \sim s$, and $b[ad] = block$, then

1. if $s \to s'$ and $b : \gamma \leadsto^n \gamma'$ then
   $((\gamma' = \bot) \lor (\gamma' \sim s' \land \forall a. q_m(a) \Rightarrow s'.m(a) = s.m(a)))$, and
2. for every $n' < n$, if $b : \gamma \leadsto^{n'} \gamma''$ then $\neg start\text{-}block(\gamma'')$.

The theorem shows (1) that if the complete execution of the block succeeds then it behaves equivalently to the ARM instruction and memory in $q_m$ is not modified, and (2) that completing the block requires exactly $n$ steps.

### 4.3 Transpiling programs

The theorems generated for every instruction are composed to verify Theorem 1. Property (1) is verified by induction over $n$, using the predicate $q_{mem}(a) \triangleq a \in \{ad \mid (ad, i) \in \pi\}$. This ensures that the ARM program is in memory after the execution of each instruction, thus allowing to make the precondition of the translation theorem (i.e. $\forall (ad_j, i_j) \in \pi. \text{read}_{32}(s.m, ad_j) = i_j$) an invariant.

Property (2) is verified by induction over $n'$. We split the execution of $n'$ steps (leading from the initial state to $\gamma'$) in two parts: $n'_0 < n'$ steps from the initial state to the last state $\gamma_0$ satisfying *start-block* and $n'_1 = n' - n'_0$ steps from $\gamma_0$ to $\gamma'$. By inductive hypothesis there must exists $n_0$ such that the ARM program reaches a state $\gamma_0 \sim s_0$ in $n_0$ steps. Since $\gamma_0 \sim s_0$ then the program counter of $\gamma_0$ points to one of the blocks produced by the transpiler. If $\gamma'$ satisfies *start-block* then we can use the corresponding instruction-theorem to show that $n'_1$ is equal to the length of the block. This and the fact that the ARM transition relation is total enables part (1) of the instruction-theorem, showing that the ARM instruction behaves equivalently to the BIL block.

### 4.4 Support for more architectures

In the following, we review the modifications of the certifying procedures needed to support other common computer architectures, like MIPS, x86 and ARMv7.

The transpiler has three main dependencies: A formal model of the architecture, a function producing step theorems, and the definitions of a simulation relation. There exist HOL4 models for x86, x64, ARMv7-M, and MIPS that are equipped with the corresponding step function. On the other hand, the simulation relation can differ for each architecture since it maps machine state fields to BIL variables. In fact, the name, the number, and the type of registers can be very different among unrelated architectures.

The expression translation has to handle the expressions of guard conditions and transformation functions that are present in the step theorems. Since these use HOL4 number and word theories, independently of the architecture, big parts of the translation of Section 4.1 can be reused. There are two exceptions: One is the possible usage different word lengths, and the other is the need of proving helper theorems to justify the emulation of operators that have no direct correspondence in BIL (e.g. for the computation of carry flag in ARM).

The transpilation of single instructions of Section 4.2 would produce BIL blocks with a similar structure. However, the changed simulation relation can affect the transpilation procedure. In fact, the BIL variables that have to be temporarily saved and the ones that must be modified can be different, matching the different registers. On the other hand, the expressions computing the state transformation are the result of the expression translation and do not require changes. Also, a jump instruction must terminate the instruction block to steer the control flow dependent on program counter changes.

The verification of Theorem 1 by the program transpilation of Section 4.3 involves only reasoning on BIL and the theorems generated for the individual instructions. This reasoning can be largely reused since the structure of these individual theorems is unchanged. Even though our proof procedure for this is fairly general, differences in the simulation relation might require slight changes.

## 5  Using the transpiler to verify binary programs

The output of the transpiler can be used to verify properties of the translated ARM program. The verification work flow consists of three tasks, (1) proving that the BIL program does not reach error states, (2) proving that the desired properties of the BIL program hold, and (3) using the refinement relation to transfer these properties to the original ARM program. Here, we show that the transpiler output fulfills this purpose for four common verification tasks: Control Flow Graph (CFG) analysis, contract-based verification, partial correctness refinement, and verification of termination.

Program's CFG is essential to many compiler optimizations and static analysis tools. Furthermore, proving control flow integrity ensures resiliency against return-oriented programming [23] and jump-oriented programming attacks [4]. In its simplest form, the CFG consists of a directed connected graph $G$, whose node set is $\mathbb{B}^{64}$, and a root node $ad_0$: The graph $G$ contains $(ad_1, ad_2)$ if the program can flow from the address $ad_1$ to the address $ad_2$ by executing a single instruction; The root node represents the entry point of the program.

Analyzing the CFG of a binary program requires to deal with indirect jumps. Even if the source program avoids using function pointers, indirect jumps are introduced by the compiler, e.g. to handle function exits and exceptions. For instance, the ARM link register is used to track the return address of functions and can be pushed to and popped from the stack. For this reason, the correctness of the control flow depends on the integrity of the stack itself. Thus, verifying the CFG $(G, ad_0)$ of a program $\pi$ requires assuming a precondition $P$, which constraints the content of the heap, stack and registers.

**Definition (Control flow graph integrity).** For every ARM state $s$ such that $stored(s, \pi)$, $s.sr.pc = ad_0$, and $P(s)$, for every $n$, if $s \to^n s_1$ and $s_1 \to s_2$ then $(s_1.sr.pc, s_2.sr.pc) \in G$.

It is straightforward to show that CFG integrity can be verified using the transpiler theorem, by defining a BIL precondition $P'$ that corresponds to $P$, and by proving the following verification conditions.

**Condition (BIL control flow integrity).** Let $lbl(\gamma) = pc$ be the label of the program counter of the state $\gamma$, which is undefined when $pc = \bot$. For every $\gamma$ such that $P'(\gamma)$ and for every $n_1$ and $n_2$, if $b : \gamma \leadsto^{n_1} \gamma_1 \leadsto^{n_2} \gamma_2$, $\textit{start-block}(\gamma_1)$ and $\textit{start-block}(\gamma_2)$, and $(\forall n_3 < n_2.b : \gamma_1 \leadsto^{n_3} \gamma_3 \Rightarrow \neg\textit{start-block}(\gamma_3))$ then $\gamma_1 \neq \bot$, $\gamma_2 \neq \bot$, and $(lbl(\gamma_1), lbl(\gamma_2)) \in G$.

**Condition (Transfer of precondition).** For every $\gamma$, $s$ such that $\gamma \sim s$, if $P(s)$ then $P'(\gamma)$.

Contract based verification consists in verification of Hoare triples to establish partial correctness. Let $P(s)$ and $Q(s, s')$ be two predicates, representing the pre- and post-condition of a contract, verifying that a program $\pi$ (starting from the entry point $ad_0$) meets the contract $(P, Q)$ means establishing the following property.

**Definition (Contract verification).** For every $s$ such that $stored(s, \pi)$, $s.sr.pc = ad_0$ and $P(s)$, for every $n_1$, if $s \to^{n_1} s_1$ then $Q(s, s_1)$.

Let $PC_{end}$ be the set of exit points of the program and $End(s_1)$ be $s_1.sr.pc \in PC_{end}$. Usually $Q$ has the form $End(s_1) \Rightarrow Q_1(s, s_1)$, meaning that if the program reached one of its exit points then the post-condition $Q_1$ is satisfied. This property can be verified using the theorem produced by the transpiler, by identifying a BIL contract $(P', Q')$, and by proving the following verification conditions:

**Condition (BIL contract verification).** For every $\gamma$ such that $P'(\gamma)$ and for every $n$, if $b : \gamma \leadsto^n \gamma'$ then $\gamma' \neq \bot$ and $Q'(\gamma, \gamma')$.

**Condition (Transfer of contracts).** For every $\gamma$, $\gamma'$, $s$, $s'$ such that $\gamma \sim s$ and $\gamma' \sim s'$, if $P(s)$ then $P'(\gamma)$ and if $Q'(\gamma, \gamma')$ then $Q(s, s')$.

Partial correctness is proved as a refinement using an abstract specification and reusing contract verification. With composability of specifications in mind, we assume that the specification is phrased such that domain and codomain are the same. Let $a_{out} = f_{spec}(a_{in})$ be a functional specification with the signature $f_{spec} : A \to A$.

**Definition (Partial correctness refinement).** For every $s, a$ such that $R(s, a)$, forall $n_1$ such that $s \to^{n_1} s_1$, if $End(s_1)$ then $R(s_1, f_{spec}(a))$.

Notice, that the refinement relation $R(s, a)$ implicitly contains the mapping from $a$ to $s$ and an invariant to enable establishing the refinement. By using the assumption $R(s, a)$, we can simply derive a verification condition in the shape of the definition for contract-based verification, which can be proved as described before. We call this the binary correctness condition in this context, where $P(s)$ resembles the invariant of the refinement relation, and $Q_1(s, s_1)$ incorporates the functional specification $f_{spec}$ with respect to the mapping of $R$.

The assertion of total correctness (or functional correctness) additionally requires termination. Therefore, we consider the following definition, where the precondition $P$ should be not stronger than the precondition we used for partial correctness (i.e. the invariant of the refinement relation).

**Definition (Termination verification).** For every $s$ such that $stored(s, \pi)$, $s.sr.pc = ad_0$ and $P(s)$, exists an $n_1$ such that $s \rightarrow^{n_1} s_1$ and $End(s_1)$.

To prove this property, we use the theorem produced by the transpiler (i.e. the second clause of Theorem 1), identify an appropriate BIL precondition $P'$, and prove the following conditions.

**Condition (BIL termination verification).** For every $\gamma$ such that $P'(\gamma)$, exists an $n$ such that $b : \gamma \rightsquigarrow^n \gamma'$ and $End'(\gamma)$.

**Condition (Transfer of termination conditions).** For every $\gamma$, $s$ such that $\gamma \sim s$, if $P(s)$ then $P'(\gamma)$ and if $End'(\gamma)$ then $End(s)$.

### 5.1 Evaluation

Our contribution counts ~4600 lines of HOL4 code: (1) ~1000 lines for the syntax and the semantics of BIL, the most of which are for the (signed an unsigned) cast operators between bitvectors of different size; (2) ~2000 for the expression transpiler, a fourth of which proves the theorems handling arithmetic conversions; (3) ~1500 for the instruction transpiler, one third of which generates the BIL fragments, and the remaining two thirds generate the proofs of correctness; and (4) the remainder for merging the instruction theorems together and generate the translation of a complete ARM program.

The whole proof-producing transpilation of an instruction takes ~ 9 s on a modern computer (Intel Core i7-6650U 2.2GHz). We follow a backward-proof strategy; firstly, we generate the proof goal by invoking the step function, merging its output, translating the expressions and generating the supposedly corresponding BIL code. This first part takes ~ 1 s. The second part symbolically evaluates the BIL statements. This takes ~ 6.5 s, with each BIL statement requiring between ~ 0.5 s and ~ 1.5 s. In the third and last part, which takes ~ 2 s, we prove that the simulation relation is established.

As described in Section 4.2, the translation of one instruction follows two steps: (i) it translates the ARM instruction to a BIL block, establishing several intermediate theorems (i.e. for translation of expressions), and setting up the goal of Theorem 2, (ii) it demonstrates Theorem 2 in a backward proof, by symbolically evaluating the BIL block and by using the intermediate theorems. The usage of a backward-proof for this procedure provides a naive strategy to speed up analyses: the user can rely on the goal produced in step (1) to translate the ARM program to BIL without generating the corresponding certification theorem. This certificate can be generated offline later. Step (2) can be optimized with additional engineering effort by using a forward-proof strategy. Furthermore, program independent helper theorems can be verified once and reused in this process.

We experimented with the transpiler using various unmodified binary programs produced by a standard GCC, including a bignum library and an implementation of AES encryption. The three C functions `internal_mul`, `newbn`, and `freebn` of the bignum library consist of 38 lines of C code, which are compiled to 141 instructions. After transpilation, we obtain 907 lines of BIL code for these functions. The encryption function of AES consists of 131 lines of C code (excluding the constant lookup tables used for the S-Boxes), which are compiled to 535 instructions. With this example, we obtain 3920 lines of BIL code. We observe that the average binary instruction consists of 6 to 7 BIL statements.

## 6  Concluding remarks

We presented the HOL4 formal model of the intermediate language of BAP and the implementation of a transpiler for ARMv8 programs. This is the first work toward this approach, and its results overcome two of the main barriers in adopting binary analysis platforms to formally verifying binary code: the lack of a formal ground to prove analysis correctness and the need for trusting translation soundness.

The formal model of BIL can be used for verifying BAP tools, which are ISA independent and analyze BIL programs, e.g., Dijkstra's weakest precondition propagation, transformation to single static assignment, loop unrolling.

In this paper, we focus on the ARMv8 architecture. To handle other machine architectures (e.g. x86, x64, ARMv7-M, MIPS), new transpilers must be developed. Fortunately, the majority of the transpiler code does not depend on specific ARMv8 features, but on the theorems produced by the step function. There are several other HOL4 models for the main commodity architectures that are equipped with the same functionality [10]. We comment on the required transpiler modifications to support these architectures in Section 4.4.

Further research is needed to develop a complete trustworthy binary analysis platform. For example, a trustworthy semi-automatic verification tool based on pre/post conditions for binary code can be implemented by completing two additional tasks: (i) a trustworthy verification condition generator to compute the weakest precondition needed by the BIL program to meet the postcondition, and (ii) a sound satisfiability solver for bitvectors to check if the precondition entails the weakest precondition. For the first task, Vogels et al. [26] verified the soundness of an algorithm for weakest precondition generation in Coq. For the second task, Satisfiability Modulo Theory (SMT) solvers can be used. Böhme et al. [5] demonstrated HOL4 proof reconstruction for Z3 [18] capable of handling the theory of fixed-size bit-vectors.

### Acknowledgments

Partially funded by framework grant "IT 2010" from the Swedish Foundation for Strategic Research, and by the KTH CERCES Center for Resilient Critical Infrastructures, which is supported by the Swedish Civil Contingencies Agency.